\PassOptionsToPackage{table,usenames,dvipsnames}{xcolor}

\documentclass[%
aps,
reprint,
 amsmath,amssymb,
prper,
floatfix
]{revtex4-2}

\usepackage{multirow}
\usepackage{graphicx}
\usepackage{dcolumn}
\usepackage{bm}
\usepackage{tikz}
\usepackage{pgfplots}
\usepackage[export]{adjustbox}
\pgfplotsset{compat=1.17}
\usepackage{listings}
\begin{document}


\title{The Boiling-Frog Problem of Physics Education}

\author{Gerd Kortemeyer}
 \email{kgerd@ethz.ch}
 \affiliation{%
Rectorate and ETH AI Center, ETH Zurich, 8092 Zurich, Switzerland
}%
\altaffiliation[also at ]{Michigan State University, East Lansing, MI 48823, USA}

\date{\today}

\begin{abstract}
It is astonishing how rapidly general-purpose AI has crossed familiar thresholds in introductory physics. Comparing outputs from successive models, GPT-5~Thinking moves far beyond the plug-and-chug tendencies seen earlier: on a classic elevator problem it works symbolically, notes when variables cancel, and verifies results; attempts to prompt novice-like behavior mainly affect tone, not method. On representation translation, the model scores 24/26 (92.3\%) on TUG-Kv4.0. In a card-sorting proxy using two of my comprehensive finals (60 items), its categories reflect solution method rather than surface features. Solving those same exams, it attains 27/30 and 25/30, with most misses in ruler-based ray tracing and circuit interpretation. On epistemology, five independent CLASS runs yield 100\% favorable, indicating a simulated expert-like stance. Framed as a ``boiling frog'' problem, the paper argues for a decisive jump: retire credit-bearing unsupervised closed-response online assessments; grade process evidence; use paper, whiteboarding; shift weight to modeling, data, and authentic labs; require transparent, citable AI use; rebuild problem types; and lean on research-based instruction and peer learning. The opportunity is to foreground what AI cannot substitute for: modeling the world, arguing from evidence, and making principled approximations.
\end{abstract}

\maketitle

\section{Introduction}
When I first evaluated GPT-3.5 in the context of physics teaching, the model would have barely cleared the calculus-based course I have taught for decades: weak with numbers, unreliable at interpreting graphics, and prone to the same kinds of novice-like reasoning errors~\cite{kortemeyer23ai} --- a surprisingly good, yet poor in absolute terms performance, and not yet a cause for concern. GPT-4o, by contrast, scored above the average post-instruction undergraduate on conceptual physics surveys~\cite{kortemeyer2025multilingual}; multimodality enabled basic diagram and image interpretation, but its reasoning remained far from expert. With the release of GPT-5~Thinking, the natural question is: where does its performance sit on the novice--expert continuum, and what should instructors take from that?

\section{Results}
\subsection{Plug-and-Chug}
\begin{figure}
\begin{center}
\includegraphics[width=\columnwidth]{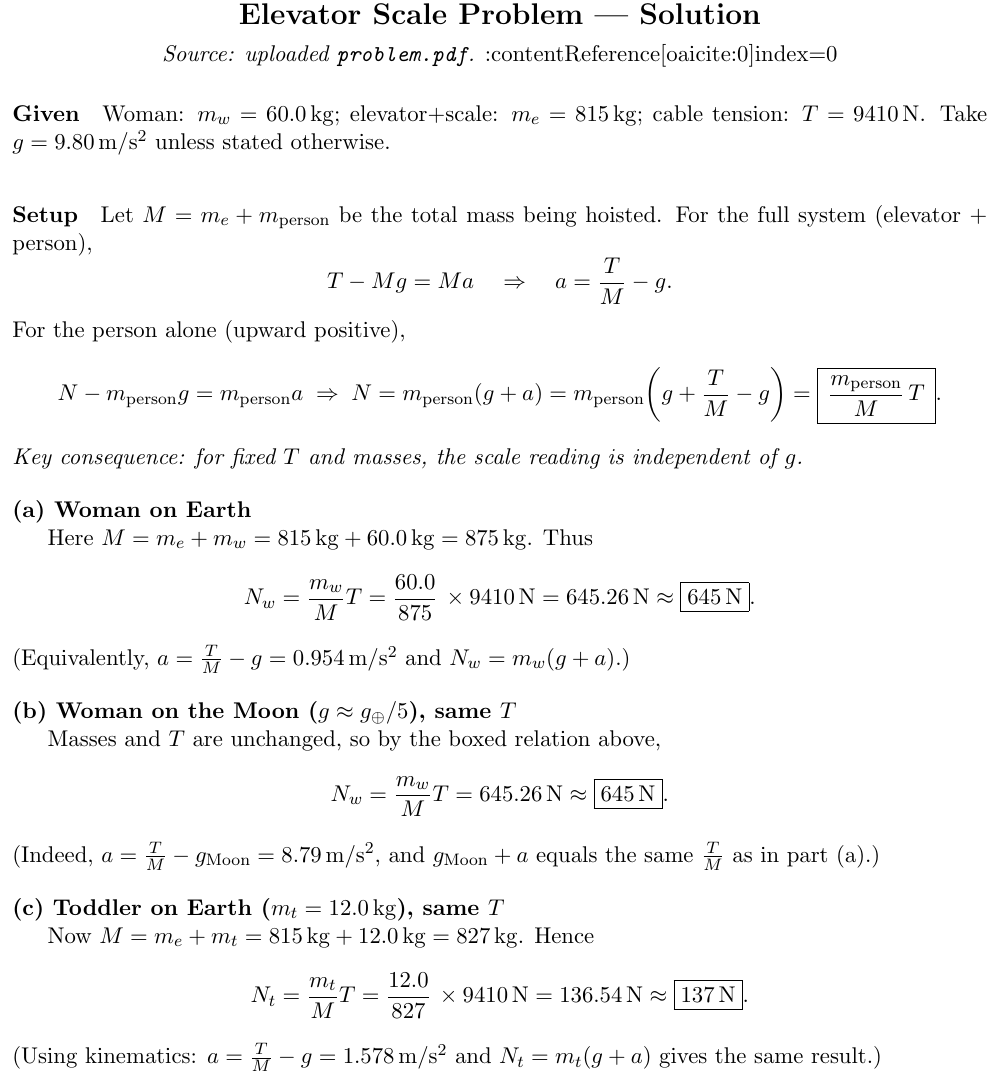}
\end{center}
\caption{The GPT-5-Thinking solution to the elevator problem.}
\label{fig:noplugchug}
\end{figure}

A decade ago, I published a mildly exasperated study on our losing battle against plug-and-chug~\cite{kortemeyer2016bat}---the novice tendency to insert numbers into formulas immediately, carry the result into the next formula, and repeat. The focal task was a standard ``elevator problem'':
\begin{quote}\small
a) A woman stands on a scale in a moving
elevator. Her mass is 60.0 kg, and the
combined mass of the elevator and the
scale is an additional 815 kg. Starting from
rest, the elevator accelerates upward.
During the acceleration, there is a tension
of 9410 N in the hoisting cable. What does
the scale read during the acceleration (in
Newton)?

b) For no good reason, the experiment is moved to the Moon, where the gravitational
acceleration is about 1/5th that on Earth. Leaving the tension in the hoisting cable the
same, what does the scale read now?

c) Back on Earth, a toddler with a mass of 12.0 kg (i.e., 1/5th the mass of the woman)
rides the elevator, again leaving the tension in the hoisting cable the same. What does the
scale read now?
\end{quote}
The problem has two common plug-and-chug pitfalls:
\begin{itemize}
\item Spending time on part (b) without noticing that the result is independent of $g$.
\item Assuming that the answer to part (c) is simply one-fifth of the answer to part (a).
\end{itemize}
When the problem is uploaded as a PDF and prompted with the straightforward request, ``solve this problem, output as LaTeX,'' the GPT-5~Thinking solution avoids both pitfalls: it proceeds symbolically, shows that $g$ cancels, and includes a quick consistency check using elementary kinematics (see Fig.~\ref{fig:noplugchug}). One might prefer retaining $m_e + m_{\text{person}}$ rather than introducing $M$, but the work is decidedly not novice.

Intriguingly, GPT-5~Thinking appears robust to prompts intended to elicit novice-like behavior: requests such as ``act like an undergraduate student'' have little effect on the problem-solving approach, beyond shifting the exposition toward a more casual, Gen-Z--tinged register. In short, the generated solutions remain algebra-first and verification-oriented rather than plug-and-chug.

\subsection{Representation Translation}
A persistent limitation observed with GPT-4o --- a multimodal system that can, in principle, process images --- was translating between mathematical or narrative descriptions and graphs, and vice versa~\cite{polverini24,kortemeyer2025multilingual}. To probe progress, I uploaded the full English version of the Test of Understanding Graphs in Kinematics (TUG-K) v4.0~\cite{beichner1994testing} as a PDF and prompted the model to solve it and output a CSV. GPT-5~Thinking scored 24/26 (92.3\%). The two missed items (15~and~21) both required matching graphs to other graphs representing their integrals. In short: where 3.5 lacked visual modality and 4o still struggled, 5~Thinking performs nearly expertly

\subsection{Card Sorting}
One of the most influential studies on the novice-expert distinction is the card-sorting experiment by Chi et al.~\cite{chi1981}, which found that novices tend to group physics problems by surface features, whereas experts group them by underlying structure. The paradigm is notoriously difficult to replicate~\cite{wolf12,wolf2012}, but it offers a useful reference point for evaluating GPT-5~Thinking on this dimension. Because the original card set is unavailable, I used the final exams from the first- and second-semester courses I taught and co-taught in Fall2008 and Spring2009, comprising 60 multiple-choice questions (the data set is available as supplemental material). The items are randomized~\cite{kortemeyer08} and not drawn from a textbook, making it unlikely they appear in training data in this form. I uploaded the entire set in one batch with the original Chi prompt: ``sort the problems based on similarity of solution, finding categories.''

\begin{table}
\caption{Problem categories by solution method across the two final exams.}
\begin{ruledtabular}
\begin{tabular}{lp{5cm}p{1.5cm}}
Semester & Category (by solution method) & Problems\\
\hline
1st & Forces \& Newton's laws (FBD, friction, N3) & 1, 2, 3, 11, 25, 26, 27, 28 \\
1st & Kinematics \& relative motion & 8, 19 \\
1st & Energy / Work / Power (incl.\ rotational energy) & 14, 18, 22, 23, 24 \\
1st & Momentum \& collisions & 12 \\
1st & Rotation \& torque (angular kinematics/dynamics) & 15, 16, 17 \\
1st & Oscillations (SHM) & 9, 29 \\
1st & Fluids \& buoyancy (continuity/Bernoulli/Archimedes) & 6, 20, 21 \\
1st & Thermodynamics \& heat (entropy/calorimetry) & 5, 7, 10 \\
1st & Waves (basic relations) & 4 \\
1st & Units \& dimensions & 13 \\
1st & Administrative (no physics) & 30 \\
2nd & Electrostatics (fields, potential, flux) & 5, 6, 12, 17, 28 \\
2nd & Magnetostatics (forces/torque on currents) & 2, 13, 15 \\
2nd & Electromagnetic induction (Faraday/Lenz) & 14, 22 \\
2nd & Circuits (DC/AC, capacitors, RC/RLC, energy in C) & 8, 9, 10, 18, 27, 29 \\
2nd & Optics (polarization, resolution, TIR, lenses) & 1, 3, 11, 21, 23, 30 \\
2nd & Waves \& EM radiation (intensity/pressure) & 19, 20 \\
2nd & Modern/relativistic/nuclear (photoelectric, decay, etc.) & 4, 16, 24, 25, 26 \\
\end{tabular}
\end{ruledtabular}
\caption{Problem categories by solution method across the two final exams.}
\label{tab:problem-categories}
\end{table}

Table~\ref{tab:problem-categories} shows the result. By the coding rules of the Chi paradigm, valid categories are neither surface-level (car problem,'' read-a-graph problem,'' etc.) nor so abstract as to collapse distinctions (``energy conservation,'' regardless of whether the context is mechanics or E\&M). The clusters produced by GPT-5~Thinking fall squarely in between: topic-oriented and solution-relevant rather than surface-driven. In short, the outcome is not novice-like.

\subsection{Final Exam Performance}
I also asked GPT-5~Thinking to solve the two exams and output a CSV. It first returned answers for items that did not require interpreting graphs, diagrams, or plots; the remaining answers followed after I uploaded screenshots of the relevant figures. The interface additionally highlighted which regions of the pages and images were being referenced --- an eye-tracking-like trace. On the first-semester exam the score was 27/30 (under the course rules, where 28 counts as 100\%, this is 96.4\%); on the second-semester exam, the score was 25/30 (89.3\% on the same scale). Errors clustered on items requiring interpretation of circuit diagrams and on questions where students were expected to construct ray diagrams with a ruler. In any case, this is a far cry from barely making the cut two years ago.

\subsection{Epistemology}
Of course, it makes little sense to ask what a language model ``believes''; the more interesting question is which beliefs, attitudes, and expectations it \emph{simulates}. Using the Colorado Learning Attitudes about Science Survey (CLASS)~\cite{adams04} via PhysPort~\cite{mckagan2020physport}, I prompted GPT-5~Thinking to complete the instrument five independent times. Each run took about a minute and yielded 100\% favorable responses. By CLASS scoring conventions, the simulated stance is entirely expert-like. Again, to be clear: CLASS is designed and validated for human respondents; this probes how the model maps items to learned response patterns, not any internal beliefs.

\section{Discussion}
In two years, the GPT series moved from D-student to A-student. Each time we said, ``at least it cannot do \emph{this} or \emph{that}'' --- calculate reliably, read graphs --- the limitation vanished within months. It still struggles with ray-tracing items that expect ruler-accurate constructions, but even that feels provisional. Even if the rate of improvement eventually flattens, much of what we emphasize in first-year physics risks feeling obsolete to students. A decade ago, asking for the hand computation of $\sqrt{54{,}031}$ would have been received as, at best, an eccentric one-off challenge and, at worst, irrelevant grunt work --- everyone already carried a supercomputer in their pocket. Many run-of-the-mill end-of-chapter problems now read the same way. For better or worse, general-purpose AI is becoming as ubiquitous as smartphones.

The well-known (and fortunately biologically inaccurate) ``boiling frog'' fable says that a frog placed in water heated gradually will not notice the danger and will fail to jump out. Its value here is purely metaphorical: incremental gains in model capabilities are easy to normalize away, until we find our assessments and learning goals simmering in a pot designed for a different era. The responsible move for physics education is to \emph{jump}: to make discontinuous changes rather than tinkering at the margins.

In my view, here are places where a jump out of the pot is warranted:
\begin{description}
\item[Online assessments for credit] Closed- or short-answer online homework, quizzes, and exams conducted outside controlled classroom settings are no longer viable sources of credit or effective gatekeepers. Rules against using AI are unenforceable; ultimately, conscientious students are disadvantaged and may feel penalized for their integrity. AI-detection tools are snake oil, and requiring lockdown browsers, cameras, keystroke loggers, microphones, etc.\ in students' homes is not only an invasion of privacy but also can look like a desperate attempt to defend an irrelevant practice, as well as  one that distracts from learning or from any sense of having fun with physics.

\item[Focus on the process] Since generative models can supply final answers, require \emph{accountable work}: modeling assumptions, unit analysis, limiting cases, estimation, and a brief ``plan,$\to$,solve,$\to$,check'' trace. Grade the reasoning, not merely the result, and award credit for productive revisions.

\item[Use paper and pencil] As you foreground process, have students work on paper, starting from a blank sheet. For feedback and grading support, AI-assisted workflows can help~\cite{kortemeyer2024grading,kortemeyer2025assessing,chen2025grading}; better yet, incorporate structured peer evaluation.

\item[Make AI use explicit and citable] Do not pretend AI is absent; require disclosure. Ask students to include representative prompts and outputs in an appendix, then \emph{critique} them: What is correct? What is spurious? How was the output verified or improved? And, before you ask: yes, GPT-5~Thinking assisted in polishing this opinion piece (grammar, flow, structure).

\item[Whiteboarding] Replace part of auto-graded credit with 3--5 minute oral spot checks (in person or synchronous online), socratic dialogue, and collaborative problem solving. This scales well in studios and recitations.

\item[Laboratories, modeling, data, and design] Emphasize experimental design, especially authentic, mobile-lab experiments (e.g., using smartphones, or even combining this with AI~\cite{vogt2025iphysicslabs}): uncertainty analysis, data cleaning, and model-to-measurement comparison. The reality of messy data and instrument idiosyncrasies is an excellent antidote to artificial answers.

\item[Rebuild problem types] Favor Fermi estimates, novel contexts with explicit assumptions, multi-representation translation (diagram $\leftrightarrow$ math $\leftrightarrow$ prose), and ``diagnose-and-repair'' tasks where students improve a flawed solution.

\item[Assessment architecture] Move major certification to supervised settings (studios, practicums, labs) and to open-resource take-home assessments that demand process evidence. Spend less energy on an arms race in proctoring and more on tasks for which unearned copying is unhelpful.

\item[Have stamina] Have the stamina to stick with research-based instructional strategies~\cite{henderson2012b}. These approaches, aimed at conceptual understanding and retention rather than mechanical mastery of algorithmic procedures, are more relevant than ever as AI improves at the latter.

\item[Ways-of-knowing and creating] Science is a human activity, and the history of physics beyond little ``history boxes'' in textbooks can be a way to exploring how knowledge is created~\cite{kortemeyer2009history}.

\item[Peer teaching and learning] Have students learn with and from each other. Not only does one learn by teaching~\cite{crouch2001peer}, but many students today are at risk of loneliness and isolation. As AI becomes increasingly ``perfect'' at routine work, it is time for humans to lean into the human parts.
\end{description}

The good news is that none of this diminishes physics; it foregrounds what makes the discipline valuable: modeling the world, arguing from evidence, and making principled approximations. With a deliberate jump now, first-year courses can become \emph{more} authentic, humane, and rigorous --- not despite AI, but because we chose to teach what AI cannot substitute for.
\bibliography{BoilingFrog}

\end{document}